\documentclass[conference]{IEEEtran}

\usepackage[T1]{fontenc}
\usepackage{multirow}
\usepackage{balance}

\usepackage[absolute]{textpos}
\usepackage[hidelinks]{hyperref}

\begin{document}

\begin{textblock}{15}(0.5,14.9)
{
\noindent\hrulefill

\noindent\fontsize{8pt}{8pt}\selectfont\copyright\ 2017 IEEE. Personal use of this material is permitted. Permission from IEEE must be obtained for all other uses, in any current or future media, including reprinting/republishing this material for advertising or promotional purposes, creating new collective works, for resale or redistribution to servers or lists, or reuse of any copyrighted component of this work in other works. \hspace{5pt} This is the accepted version of: M. Sul\'ir, J. Porub\"an, O. Zori\v{c}\'ak. IDE-Independent Program Comprehension Tools via Source File Overwriting. 2017 IEEE 14th International Scientific Conference on Informatics, IEEE, 2017, pp. 372--376. \url{http://doi.org/10.1109/INFORMATICS.2017.8327277}

}
\end{textblock}

\title{IDE-Independent Program Comprehension Tools\\via Source File Overwriting}

\author{\IEEEauthorblockN{Mat\'u\v{s} Sul\'ir, Jaroslav Porub\"an and Ondrej Zori\v{c}\'ak}
\IEEEauthorblockA{Department of Computers and Informatics\\
Faculty of Electrical Engineering and Informatics\\
Technical University of Ko\v{s}ice\\
Letn\'a 9, 042 00 Ko\v{s}ice, Slovakia\\
Email: \{matus.sulir,jaroslav.poruban\}@tuke.sk,\\ondrej.zoricak.2@student.tuke.sk}
}

\maketitle

\begin{abstract}
Traditionally, we have two possibilities to design tools for program comprehension and analysis. The first option is to create a standalone program, independent of any source code editor. This way, the act of source code editing is separated from the act of viewing the code analysis results. The second option is to create a plugin for a specific IDE (integrated development environment) -- in this case, a separate version must be created for each IDE. We propose an approach where information about source code elements is written directly into source files as annotations or special comments. Before committing to a version control system, the annotations are removed from the source code to avoid code pollution. We briefly evaluate the approach and delineate its limitations.
\end{abstract}

\IEEEpeerreviewmaketitle

\section{Introduction}

A programming language should serve as a mean of communication between a human and a computer. The valid source code of a program, by definition, contains all required information to be unambiguously compiled and executed on a computer. On the other hand, the source code is often understood by humans with great difficulties, or not at all. This can be attributed to a lack of information useful for humans contained in the code.

To improve program comprehension, a vast variety of tools analyze the source code, a running program, the programmer's interaction, the version control system and other external systems. Then, they associate the obtained metadata with parts of the source code. In our article \cite{Sulir17labeling}, we performed a systematic mapping study of such approaches and tools. We surveyed existing literature related to labeling (augmenting) the source code and categorized the approaches into a taxonomy with four dimensions: source, target, presentation and persistence.

To clarify the mean of source code labeling, let us illustrate it with specific examples. Stacksplorer \cite{Karrer11stacksplorer} statically analyzes the source code and visually annotates method definitions in the IDE with lists of their callers. Senseo \cite{Rothlisberger12exploiting} displays method callers obtained using dynamic analysis; in addition, it shows metrics like execution frequencies or memory consumption of methods. IDE plugins such as Deep Intellisense \cite{Holmes08deep} and Rationalizer \cite{Bradley11supporting} show metadata like an author, time or message of the last commit next to each source code line.

\subsection{Motivation}

Currently, when the aim of a third-party tool is to label parts of the source code with metadata, the tool authors have two main implementation possibilities:
\begin{itemize}
\item an IDE (integrated development environment) plugin
\item or a standalone tool.
\end{itemize}
Each of these two possibilities has its advantages and shortcomings.

IDE plugins allow for the utilization of a tight integration between the source code, the current IDE features and the new feature of the tool. For example, we can represent a comprehensibility metric with a light or dark red background directly in the editable source code view of the IDE \cite{Beyer10depdigger}. The background color supplements the existing source code highlighting and the programmer can view this augmentation and edit the source code at the same time, without switching between two separate tools.

However, plugins have also disadvantages. Since individual IDEs have very different plugin development interfaces, large parts of plugins must be developed individually for each supported IDE. For example, to support only some of the most popular Java IDEs, we must create a separate plugin for Eclipse, IntelliJ IDEA, NetBeans and JDeveloper. The situation becomes even worse when we have to consider also text editors like jEdit, Vim and Emacs. Finally, there are slight differences between individual versions of the same IDE, so a plugin developed for Eclipse 4.7 may be incompatible with Eclipse 4.3 and vice versa.

The second possibility is to create a standalone tool which displays the source code and metadata separately, without any connection with an IDE. This poses a cognitive burden on the developer -- he must switch back and forth between two tools, navigate to correct parts in both of them and mentally connect the corresponding parts. This is called a split-attention effect \cite{Beck13in}.

On the other hand, the advantage of separate tools is a lower creation and maintenance cost for the tool author.

\subsection{Synopsis}

In this paper, we would like to present an approach which does not require a creation of a separate plugin for each IDE, while at the same time, the metadata will be displayed near related source code parts and directly in the IDE. This can be achieved by tools writing the metadata directly in the source code files. We will limit the metadata only to textual information and write it in a form of Java annotations or source code comments next to the corresponding source code elements or lines. When the programmer will no longer need them, or when the source code will be about to be committed into a version control system, the metadata will be removed from the source code.

In our previous work \cite{Sulir17exposing}, we described an approach where run-time metadata (obtained using dynamic analysis) were written directly into source code using Java annotations. We distinguished between two operation modes:
\begin{itemize}
\item local-only workflow
\item and shared workflow (when the metadata were not discarded, but committed into a version control system).
\end{itemize}

The local-only workflow is similar to the approach described in this article. However, we extend it:
\begin{itemize}
\item to support various kinds of metadata, not only information obtained using dynamic analysis
\item and to support also line-level granularity by using source code comments.
\end{itemize}

Additionally, we will describe the approach in more detail, especially from the point of view of individual IDEs. Finally, we also provide a preliminary implementation of the approach, which was not available so far.

\section{Source File Overwriting Approach}

Now we will conceptually describe our approach in a high-level manner. The workflow begins with the clean source code of a program, without any metadata present.

\subsection{Annotation-Based Metadata}

Let us consider a situation when a programmer wants to analyze the program using a tool which outputs results for each class, method, member variable or another annotable element in the source code. He runs a tool, independent of a particular IDE. The tool overwrites the existing source code files, writing an annotation above each relevant element. The resulting source code has this form (Java syntax):

\begin{verbatim}
@Metadata{value}
source_code_element
\end{verbatim}

The \verb|@Metadata| annotation must be unique to the given type of analysis and not present in the clean source code.

For example, if a tool produces a numeric metric for each method, a hypothetical source code snippet looks like this:

\begin{verbatim}
@SomeMetric{1.7}
public void method() {
  ...
}
\end{verbatim}

Suppose the developer inspected the source code along with the metrics and modified it with respect to some task. Now he no longer needs the metrics to be displayed -- for example, he starts to work on another task, or he wants to commit his changes to a version control system (VCS). There are two possibilities:

\begin{itemize}
\item He manually executes the tool, instructing it to remove the annotations from the source code (e.g., by command-line arguments).
\item The tool is running in the background, waiting for the VCS command to start. When it starts, the tool pauses the VCS process, removes the annotations automatically and then resumes the process.
\end{itemize}

Either way, the source code submitted to a VCS will be clean and co-workers will not be aware of the programmer using the tool.

\subsection{Comment-Based Metadata}

Sometimes, method-, class- or member-level granularity is not enough. In case we need to assign line-level metadata, we can use one-line source code comments instead of annotations. To ensure the comments can be unambiguously removed after they are no longer necessary, they must have a specific format, not present in the clean, unannotated code. In general, lines labeled with metadata have this form:

\begin{verbatim}
code_line //special_character(s) metadata
\end{verbatim}

For instance, suppose we want to display code coverage results. We need to assign a piece of metadata to each executable line of code: whether a line was executed or not. Then a possible source code snippet can look like this:

\begin{verbatim}
public void method() {
  int i = 1; //+ executed
  if (i == 2) { //+ executed
    doSomething(); //+ not executed
  }
}
\end{verbatim}

After the metadata are no longer needed by the programmer, they can be removed in the same way as in the case of Java annotations: either manually or as an automatic reaction to a VCS process execution.

\section{Tool Description}

A preliminary implementation of our approach is named SOverwrite. The source code of the tool is available online\footnote{http://bit.do/SOverwrite}.

\subsection{Possibilities}

The current version of SOverwrite includes three sample analysis types to demonstrate our approach:

\begin{itemize}
\item static analysis: callers of methods,
\item dynamic analysis: code coverage,
\item version control system: authors of last class/method modifications.
\end{itemize}

\subsubsection{Method Callers}

The ``method callers'' sub-tool statically analyses the source code and inserts  the Java annotation named \verb|@Caller| above each method which has exactly one caller, for example:

\begin{verbatim}
@Caller{clazz=Math.class, method="square"}
public void multiply(int a, int b) {
  ...
}
\end{verbatim}

If a method has multiple callers, they are grouped using the \verb|@Callers| annotation:

\begin{verbatim}
@Callers{
 @Caller{clazz=Mat.class, method="square"}
 @Caller{clazz=Mat.class, method="fact"}
}
public void multiply(int a, int b) {
  ...
}
\end{verbatim}

\subsubsection{Code Coverage}

The ``code coverage'' sub-tool executes the program and counts how many times each line was executed (using an instrumentation agent). To each line which was executed at least once, it appends a specially-formatted comment representing the execution count of this particular line. A sample snippet looks like this:

\begin{verbatim}
public void method() {
  int i = 0; //+1
  while (i < 2) { //+3
    doSomething(); //+2
    i++; //+2
  }
}
\end{verbatim}

In the example, the second line was executed once, the third three times, etc.

\subsubsection{Last Author}

The ``last author'' sub-tool reads the version control system information (in our case, Git). Above each class and method, it inserts the Java annotation \verb|Author| containing a name of the last author who modified any line of the class/method. An example follows.

\begin{verbatim}
@Author("John Doe")
public void method() {
  ...
}
\end{verbatim}

\subsection{Tool Usage}

SOverwrite contains a set of configuration files which can be used for customization.

First, the programmer configures the location of the project to work with (its root directory). If some kind of dynamic analysis is going to be performed, a run-time configuration, such as the location of the executable and command-line arguments, is also necessary.

The user can also select which types of analysis (sub-tools) will be active. The analysis can be run on the whole source code tree or just for the selected parts: It is possible to select classes and methods to be labeled with metadata, while the other ones will remain untouched.

Then, the developer executes SOverwrite to perform the selected analysis kinds and label the source with desired metadata.

Finally, the tool currently does not support the automated removal of metadata before a VCS commit. Therefore, the developer manually triggers the removal command in SOverwrite when he considers it is necessary.

\subsection{Implementation Details}

Each analysis kind is implemented as a subclass of the class \texttt{AnalyseProcessor}. Therefore, it is possible to add new sub-tools to SOverwrite. When practical, the analysis is performed in parallel on multiple files to speed up the process.

To obtain VCS information, we used the JGit\footnote{https://www.eclipse.org/jgit/} library. For static analysis and source code modification, the libraries Roaster\footnote{https://github.com/forge/roaster} and Spoon \cite{Pawlak16spoon} were used. Finally, dynamic analysis is possible thanks to the Java agent instrumentation API (application programming interface).

\section{Evaluation}

A brief evaluation of SOverwrite regarding the unwanted differences in files and automatic file reloading in IDEs follows.

\subsection{Differences in Files}

If the metadata are written into the source code and subsequently removed from it, all files should remain the same as before performing the analysis (suppose the developer did not modify them meanwhile).

If we call the labeling process $label$ and the removal process $remove$, the following should hold regarding the source code ($code$):
\[
remove(label(code)) = code
\]

Generally, SOverwrite fulfills this property. However, there is one exception -- if annotation-based metadata are used, the imports of annotation classes remain in the file. We consider this an implementation issue and not an inherent property of the approach.

\subsection{Automatic Reloading}

It is convenient to trigger an analysis while the project of interest is open in an IDE. Therefore, a file which is currently displayed in the IDE is modified by another process. This can possibly lead to confusion if the file is not synchronized. The user will not see the results of the analysis before closing and reopening the given file, which would limit the usability of the tool.

Fortunately, many IDEs and text editors offer a possibility to automatically reload a file if it is modified by another process. We summarized the possibilities and behavior of selected popular Java IDEs and text editors in Table~\ref{t:reload}.

\begin{table*}
\renewcommand{\arraystretch}{1.3}
\caption{File-reloading capability of selected IDEs and text editors}
\label{t:reload}
\centering
\begin{tabular}{|l|c|c|c|}
\hline
\textbf{IDE or text editor} & \textbf{Manual file reloading} & \multicolumn{2}{c|}{\textbf{Automatic file reloading}} \\ \cline{3-4}
& & supported & default \\ \hline \hline
Eclipse & supported & yes & off \\ \hline
IntelliJ IDEA & supported & yes & off \\ \hline
NetBeans & supported & yes & off \\ \hline
jEdit & supported & yes & on \\ \hline
SciTE & supported & yes & on \\ \hline
Vim & supported & yes & off \\ \hline
\end{tabular}
\end{table*}

We can see that both manual (on-request) and automatic (in background) file reloading is available in all surveyed applications. Furthermore, in some text editors, automatic file reloading is turned on by default. When the default option is off, it can be easily adjusted in the application's settings. Therefore, we can conclude that file reloading is not an issue at all.

\section{Discussion}

We will now discuss the limitations of our approach and its applicability to programming languages other than Java.

\subsection{Limitations}

The most prominent limitation of our approach is its usability only for purely textual metadata. Insertion of graphics, such as plots, diagrams, etc., is inherently not possible in its current version. However, Schugerl et al. \cite{Schugerl09beyond} present an Eclipse plugin SE-Editor which displays images directly in the IDE code editor -- in places where specially-formatted comments are inserted in the source code (e.g., \verb|/*** http://image.url */|). If these two approaches were combined and SE-Editor was implemented also for other IDEs and text editors, it would effectively mean our approach could support also graphical elements.

Another practical limitation is the non-interactivity of the approach. Standard IDE plugins offer rich interaction possibilities. In contrast to them, the inserted annotations and comments are generally static and non-navigable. However, there are some exceptions. For example, if a Java annotation contains a parameter of type ``class'' (e.g., \verb|Math.class|), it is clickable in many IDEs. For instance, in IntelliJ IDEA, clicking on a class name while holding the Ctrl key opens the given class in the IDE.

\subsection{Applicability to Other Languages}

Although SOverwrite is implemented in Java, our approach is applicable to almost every programming language.

Annotation-based labeling requires the given language to support attribute-oriented programming. Examples of such languages are Java and C\#.

Comment-based labeling requires the given language to have one-line comments which can be appended at the end of lines. The range of languages supporting one-line comments is very broad.

\section{Related Work}

Source code annotations are traditionally designed to be processed by tools automatically. For example, an annotation above a member variable can denote its correspondence to an XML element \cite{Chodarev16development}. On the other hand, concern annotations \cite{Sulir16recording} and specially formatted tags \cite{Storey09how} contain information useful mainly for humans. However, both of them are usually shared between developers and committed to a VCS. In contrast to them, our approach utilizes annotations only temporarily, in one developer's workspace.

Lee et al. \cite{Lee08towards} designed a generic infrastructure for framework-specific IDE extensions. They try to reduce authors' work when creating an IDE extension for a new object-oriented framework. However, providing an infrastructure for multiple different IDEs is out of the scope of their work.

Asenov et al. \cite{Asenov16ide} present an IDE which enables programmers to query and modify programs using a combination of source code and other resources. Custom queries can be written using scripts, and the queries are composable using pipes. While they try to clear the line between plugin authors and application developers, their approach is currently limited to a single IDE which they created from scratch. In contrast to them, we try to reuse existing IDEs as much as possible.

Juh\'ar \cite{Juhar16integrating} discusses how to integrate concerns with IDEs using concern-oriented projections. However, such approaches are usually not IDE-independent.

In \cite{Sulir15semiautomatic}, we presented a semi-automated approach AutoAnnot to assign concern annotations to methods. More specifically, we tagged methods in the source code with feature tags such as \verb|@NoteAdding| or \verb|@FileSaving|. This was achieved using software reconnaissance \cite{Wilde95software} which computes a difference of sets of executed methods between two runs. Compared to AutoAnnot, SOverwrite is much more general and supports also comment-based labeling.

As we already mentioned, Schugerl et al. \cite{Schugerl09beyond} present a plugin to display images in the source code view. This could be used to extend the possibilities of SOverwrite.

\section{Conclusion and Future Work}

In this article, we presented an approach of program comprehension and analysis tools independent of a specific IDE. While the source code is open in an IDE, it is modified by a tool -- metadata like analysis results are written directly into the code in a form of annotations or comments. Therefore, the tool is not aware of the IDE and the programmer can use an environment or text editor of his or her choice. When the metadata are no longer necessary, they are removed to prevent VCS pollution.

A sample implementation, SOverwrite, was described in the article too. We demonstrated how the source code can be labeled with the results of static and dynamic analysis, along with VCS information.

The sub-tools (analysis types) currently implemented in SOverwrite act only as a demonstration. In the future, we would like to perceive it as a framework to which tool writers can add custom features.

The system should be also optimized and performance evaluation should be performed. Assessing the impact on developers' work would be a useful future work direction.

\section*{Acknowledgment}
This work was supported by project KEGA 047TUKE-4/2016 Integrating software processes into the teaching of programming.

This work was also supported by the FEI TUKE Grant no. FEI-2015-23 Pattern based
domain-specific language development.

\bibliographystyle{IEEEtran}
\balance
\bibliography{informatics}

\end{document}